\documentstyle[aps,epsf,titlepage,amsmath,amssymb]{revtex} 
\begin{document}

\tolerance 10000

\draft

\title{Slave Bosons in Radial Gauge:\\
the Correct Functional Integral Representation and 
Inclusion of Non-Local Interactions}

\author{Raymond Fr\'esard$^{1,a}$\thanks{Email: Raymond.Fresard@ismra.fr} 
and
Thilo Kopp$^{2}$\thanks{Email: Thilo.Kopp@physik.uni-augsburg.de}}
\address{
$^1$ Institut de Physique, Universit\'e de Neuch\^atel, A.-L.\ Breguet 1, 
   2000 Neuch\^atel, Switzerland\\
$^{2}$ Experimentalphysik VI, Elektronische Korrelationen und
Magnetismus, Universit\"at 
Augsburg, D-86135 Augsburg, Germany\\[3em]}
\maketitle
\begin{abstract}
We introduce a new path integral representation for slave bosons
in the radial gauge which is valid beyond the conventional fluctuation
corrections to a mean-field solution. For electronic lattice models, defined
on the constrained Fock space with no double occupancy, all phase fluctuations
of the slave particles can be gauged away if the Lagrange 
multipliers which enforce the constraint on each lattice site are 
promoted to time-dependent fields. 
Consequently, only the amplitude (radial part) of the slave boson
fields survives. It has the special property that it is  equal to 
its square in the physical subspace. This renders the functional integral 
for the radial field Gaussian, even when non-local Coulomb-type interactions 
are included. We propose i) a continuum
integral representation for the set-up of further approximation
schemes, and ii) a discrete representation with an Ising-like radial
variable, valid for long-ranged interactions as well. The latter scheme
can be taken as a starting point for numerical evaluations.
\end{abstract}

\newpage

\section{Introduction}

\noindent The significance of strong electronic correlations in metals
has long been acknowledged when intractable enigmas of solid state 
physics such as itinerant magnetism, the correlation-induced 
metal-insulator transition, unconventional superconductivity, or 
spin liquid states and non-Fermi liquid behavior of electronic
excitations were embraced. The scope of the ``strongly
correlated physics'' is nowadays
enormous; it extends to the fractional quantum Hall effect and
high-$T_c$ superconductivity, and also in mesoscopic systems an
increasing number of investigations take this route since
certain scenarios with strong local interactions
can be modeled more directly in experiments on a mesoscopic scale. The
efforts to fathom the underlying mechanisms resulted only in partial 
success. Especially, models with strong local interactions of itinerant 
electrons resisted a thorough understanding beyond
certain limiting, and often unphysical, cases.

The celebrated Hubbard model epitomizes the two opposing characters of
narrow band electrons: electron hopping (with energy scale $t$) supports
the itinerant character, suitably expressed in momentum space, and 
favors a metallic character. In contrast to this, the local on-site Coulomb 
interaction $U$ may drive the electrons, depending on band-filling, into 
a Mott insulating state, with strong magnetic correlations. Standard
perturbation theory is not applicable in the transition
regime (for a recent discussion see \cite{Vilk}), and whether it even
was valid for any finite $U$ in two space dimensions \cite{Anderson}
was a longstanding problem. On the
other hand, perturbation theory from the atomic limit, with
no kinetic terms, is not feasible in a straightforward manner
since the tremendous degeneracy for low energy states of the
fully localized limit renders the usual perturbation theory inapplicable.
Furthermore, a hopping expansion cannot build on Wick's theorem since 
the zeroth order Hamiltonian is not bilinear in the canonical electron fields,
though progress has been achieved recently
\cite{Pairault,Kampf}. 
Nonetheless, for $ t \ll U$ a Schrieffer-Wolff
transformation may be performed to the appropriate order in $t/U$.
This may be visualized as a rotation of the states in Hilbert space
such that transitions between different subspaces with fixed number of 
doubly occupied sites are suppressed to the corresponding
order in $t/U$. Equivalently, one usually applies a canonical transformation
to the Hamilton operator in order to generate an effective Hamiltonian 
of which the low energy part is
explicit. The Hubbard model is reduced in this way \cite{Takahashi} to a $t$--$J$
model (plus pair-hopping) up to linear order in $t/U$ --- or to a
``$t$-model'' in zeroth order. The
$t$-model is just the kinetic term, however it is now a
correlated hopping in
the sense that double occupancies are excluded. This constrained
hopping can only be
envisioned in real space. The $t$--$J$ model includes, beyond the
constrained hopping, a
Heisenberg spin-exchange and a short range density-density
correlation
term of scale $J \sim t^2/U$.
Although all these strong coupling models
present the local spin and charge
excitations in an appealing way, it is still a long way to extract the
true (delocalized) low-energy excitations.

In the past decades considerable effort has been invested into the
construction of techniques which work in the constrained Fock space. 
Already Hubbard introduced projection operators onto local (site-) 
states with no electron ($|0\rangle$), one electron with spin $\sigma$ 
($|\sigma\rangle$) or double occupancy ($|2\rangle$). These
so-called Hubbard X-operators belong to a graded Lie-algebra, and a
canonical many-particle Green's function formalism is not applicable (for a
detailed review of functional integral representations and a suitable 
choice of coherent states for X-operators see ref.~\cite{Tuengler}). 
The evaluation scheme by Hubbard
shaped the appropriate picture of lower and upper
``Hubbard bands'' for low and high
energy (in-) coherent excitations, respectively. Yet it
suffered from inconsistencies,
e.g.\ the failure to satisfy the Luttinger theorem.

An alternative approach, which also implements a projection
onto local states, is the slave boson formalism
\cite{Barnes,Kotliar_R}. Electron creation and
annihilation operators are represented by
composite operators whereby an electron is annihilated either
by the creation of a hole site, accompanied by the annihilation 
of a singly occupied site with corresponding spin
or the annihilation of a doubly occupied site, accompanied by
the creation of a singly
occupied site with opposite spin. Several choices exist how
to ascribe fermionic and
bosonic degrees of freedom to these operators,
for a discussion see \cite{Fresard_W}.
Although the operators
are now canonical fermionic and bosonic operators, they live in a
constrained Fock space
since the sum of their respective particle numbers (the {\it
pseudo charge}) has to be
unity for each site. Mean field or saddle point approximations
satisfy the constraint only
on the average and, furthermore, lead to Bose condensation of the
slave bosons
\cite{Kopp}. Actually, the saddle point approximation becomes
exact in the unphysical
limit ${\cal N}\rightarrow\infty$, where ${\cal N}$ is the spin
degeneracy. Corrections to the mean
field result were calculated in the first order of the expansion
parameter $1/{\cal N}$ --- the
constraint is thereby replaced by a ``soft constraint'' with
pseudo charge equal to
${\cal N}/2$. For ${\cal N}=\infty$, the bosons condense, that is,
they attain a finite complex expectation value, the square of which is 
the density of empty sites.

In the Barnes representation, only the charge degree of freedom is
represented by the slave boson. In contrast, both spin and charge degrees
of freedom are carried by bosons (and fermions) in the Kotliar and
Ruckenstein representation \cite{Kotliar_R}, while
in the spin and charge rotation invariant representation
\cite{Fresard_W} they are carried by
bosons only. The mean-field approach to the Kotliar and
Ruckenstein representation has been quite successful 
when compared to numerical simulations. Indeed, the
paramagnetic mean-field solution is equivalent to the Gutzwiller
approximation\cite{Kotliar_R}. Even though this is the saddle point of
an action, it turns out to obey a variational principle in the limit of 
large spatial dimensions, where the Gutzwiller approximation and the 
Gutzwiller wave function are identical \cite{Metzner_V}. When the dimension is
reduced, they are not known to differ markedly. Comparison of ground state 
energies \cite{Lilly,Fres1}, and charge structure factors \cite{Fres2} with 
Quantum Monte Carlo data show excellent agreement. 
A variety of saddle points have been investigated, in particular in 
the vicinity of the Mott transition \cite{Tandon,Seibold,Fres_K,Hasegawa}.
The Mott gap can be calculated as well \cite{Lavagna}, and comparison with 
experimental data in La${}_x$Y${}_{1-x}$TiO${}_3$
\cite{Okimoto} displays a very good agreement \cite{Fres_K}. 
Nevertheless the computation of fluctuations, in particular in the 
magnetic channel, remains problematic.

Moreover such a finite expectation value of the complex slave boson
field violates Elitzur's theorem according to which, in a theory with 
local gauge invariance, only locally gauge invariant operators may 
acquire a non-zero expectation value. Correspondingly, for finite 
${\cal N}$, the phase fluctuations of the boson field suppress the
condensation. A zero expectation value of the bosonic field,
however, is problematic  as a starting point since the bosonic field 
has to connect to the number of empty sites, in
the end. It was early on recognized that these difficulties are
avoided in a polar decomposition where the zero mode of the phase
fluctuations may be singled out and the
expectation value of the radial amplitude is, in general,
non-vanishing \cite{Coleman0,Read} (a more recent,
detailed discussion of $1/{\cal N}$--corrections in the ``radial gauge''
is found in the review by
Arrigoni {\it et al.}~\cite{Arrigoni}). Although subtleties are
associated with the
measure and the correct translation of the time discretized
functional integral into the
continuum limit version, the result in $1/{\cal N}$ is not different from
that of an X-operator approach \cite{Gehlhoff}.

Despite the progress in the understanding of the fluctuations
around the mean field
saddle point, the physically most interesting case with the
largest possible spin
fluctuations, ${\cal N}=2$, remains inaccessible. Moreover, the saddle
point solution is
generically a Fermi liquid solution and, therefore, non-Fermi
liquid behavior is
necessarily non-perturbative in $1/{\cal N}$. This motivated the Karlsruhe
group \cite{Kroha} to device a slave-boson formalism which builds on
canonical many-particle techniques, is locally gauge-invariant by construction 
and, most importantly, allows for Fermi liquid and non-Fermi liquid 
behavior for the considered (multi-channel) Anderson impurity problem. 
It permits to recover the physical quantities over the complete temperature
range, from the high temperature local moment regime to the low
temperature correlated many-body state. In spite of this major success
of slave boson theory, it is still a formidable problem to set up a 
similar diagrammatic evaluation for strongly correlated itinerant electrons.

It is quite instructive to compare the implementation of the
constraint in this diagrammatic formalism with that in the functional
integral radial gauge representation. In the first approach, local
gauge invariance is
guaranteed through Ward
identities in a conserving approximation. This enforces the
constraint
only in so far as
transitions between sectors of the Fock space with different pseudo
charge are suppressed during the time evolution of a considered many
particle
state. The projection
onto the physical sector of the Fock space is achieved in a last
step,
when physical
expectation values are calculated, using the Abrikosov procedure of
sending the Lagrange
multiplier of the constraint to infinity. This may be contrasted with
the functional
integral calculation in the radial gauge: here the phase is fixed and
the amplitude of the
boson will in turn strongly fluctuate. The pseudo charge conservation
will not be
satisfied a priori. This is precisely the reason why one has to
project onto the physical
subspace at each time step (and, of course, at each site) which is
accomplished by
promoting the Lagrange multiplier of the constraint into a field.

In this paper we reexamine functional integrals for strongly
correlated electrons with
${\cal N}=2$ in the radial gauge. We restrain our considerations to
``$U\rightarrow\infty$'',
that is the limit of strong local interaction between electrons so
that the number of
doubly occupied sites is zero. The Hubbard model is reduced to either
the $t-$ or $t-J$
model. In this limit, the bosonic fields in radial representation
reduce to their
respective (real) amplitude since all time derivatives of the
corresponding phases can
be absorbed in a (time-dependent) Lagrange multiplier field.

It is possible to construct the functional integral either as an
integration over
continuous values of the boson radial amplitudes and the Lagrange
multiplier, or,
equivalently one can implement the constraint first and thereby
reduce
the integration
over the slave boson amplitude to a discrete sum over the values of
``hole occupancy'', $\{0,1\}$. The first method is more
appropriate for approximation schemes
which further manipulate the path integral analytically, and the
latter method is to be
taken as a starting point for numerical calculations in the strong
coupling limit which do not build on a Hirsch decoupling
\cite{Hirsch}.

In general, there are no ab initio advantages of one
functional integral representation
over another representation. It depends on the approximation
scheme if a concrete
representation is suitable. For example, when
$1/{\cal N}$--fluctuations are considered the radial gauge
consistently separates the zero-mode fluctuations. Moreover,
for any ${\cal N}$, the square of
the slave boson amplitude is equal to the amplitude itself
since the constraint projects the amplitude
onto the values $\{0,1\}$. In the radial gauge, this renders the
integrations over the real fields Gaussian, albeit the bosons are 
still coupled to the fermionic Grassmann fields. This projection 
property also ensures that the measure does not produce the 
cumbersome logarithmic terms in the action and, more significantly, non-local 
density-density interactions turn out to be Gaussian
terms in the functional integral. The latter is not valid in
the Cartesian gauge. These
observations should suffice as motivation to readdress the
construction of the slave
boson functional integral in the radial gauge.

In Section~\ref{sec:FI} we introduce the correct functional integral
representation in the radial gauge, in the limit $U
\rightarrow \infty$. In Section~\ref{sec:discrete} we propose an alternative
discrete representation which may serve for numerical
simulations. In Section~\ref{sec:nonloc} we extend these representations to
lattice models, with (possibly) long-ranged density-density
interactions. Several calculations for simple models are
carried out to illustrate this new method.

\section{ Functional integral representation in the radial
gauge.}\label{sec:FI}

In any slave boson representation of strongly correlated
electron systems there is a gauge symmetry group as a
consequence of the redundancy of the representation of the
physical electron
operator. The first slave boson representation is due
to Barnes \cite{Barnes}. It was introduced to tackle
the single-impurity
Anderson model. The local physical electron operator
$c^{\dagger}_{\sigma}$ is, in the limit of infinitely strong 
on-site interaction, rewritten as a product of auxiliary fields
$f^{\dagger}_{\sigma}$ and $b$:
\begin{equation}\label{Cop}
   c^{\dagger}_{\sigma} = b f^{\dagger}_{\sigma}
\end{equation}
Here $f^{\dagger}_{\sigma}$ is a doublet of canonical
fermionic fields, and $b$ a canonical bosonic field. We
stay in the physical sector of the Fock space provided
the auxiliary fields fulfill the constraint relation:
\begin{equation}\label{Eqcon}
Q \equiv \sum_{\sigma} f^{\dagger}_{\sigma} f_{\sigma} +
b^{\dagger} b
 = 1
\end{equation}
In a functional integral representation of the partition sum
$Z$, the constraint is enforced via the integration over a
Lagrange multiplier $\lambda$ (for a standard reference see
\cite{Bickers}): \begin{equation}\label{Zcan}
   Z = \int_{-\pi/\beta}^{\pi/\beta} \frac{\beta d\lambda }{2
   \pi} e^{i\beta \lambda} \int \prod_{\sigma}
   D[f_{\sigma},f^{\dagger}_{\sigma}] \int D[b,b^{\dagger}]
   e^{-\int_0^{\beta} d \tau {\mathcal{L'}}(\tau)}
\end{equation}
with
\begin{equation}
{\mathcal{L'}}(\tau) = {\mathcal{L}}_f(\tau) +
{\mathcal{L}}_b(\tau) + {\mathcal{L}}_t(\tau) +
{\mathcal{L}}_{\mathnormal{\rm nloc}}(\tau) \end{equation}
and
\begin{eqnarray}\label{Lcont}
{\mathcal{L}}_f(\tau) &=& \sum_{\sigma}
f^{\dagger}_{\sigma}(\tau) (\partial_{\tau} - \mu +
i\lambda)f_{\sigma}(\tau) \nonumber\\ {\mathcal{L}}_b(\tau)
&=& b^{\dagger}(\tau)
(\partial_{\tau} + i\lambda) b(\tau)
\end{eqnarray}
The last two contributions to the Lagrange density are
hopping of electrons (${\mathcal{L}}_t$) and the non-local
part of the electron-electron interaction
(${\mathcal{L}}_{\rm nloc}$), respectively. The chemical
potential, which fixes the average number of electrons, is
$\mu$. The space dependence (lattice site index $i$) of the
fields and of $\lambda$ is suppressed in Eq.~(\ref{Zcan}),
for
simplicity. Since the pseudo charge $Q$, defined in
Eq.~(\ref{Eqcon}), is a conserved quantity, the action is
invariant under a group of local $U(1)$ gauge 
transformations \cite{Read}. In the continuum limit, it reads:
\begin{eqnarray}\label{Eqgauge}
f_{\sigma}(\tau) &\rightarrow
\tilde{f}_{\sigma}(\tau) = & f_{\sigma}(\tau)
e^{i\varphi(\tau)}\nonumber\\
b(\tau) &\rightarrow  \tilde{b}(\tau) = &
b(\tau)e^{i\varphi(\tau)}\nonumber\\
\lambda &\rightarrow  \tilde{\lambda}(\tau) =
&\lambda - \dot{\varphi}(\tau) \quad .
\end{eqnarray}
The physical electron operators are invariant under this
transformation. It is well known that $\lambda$ has to be
continued into the complex plane, and the integration contour
has to be shifted into the lower half-plane in order to ensure
convergence of the functional integral \cite{Bickers2}:
\begin{equation}
\lambda \rightarrow \lambda -i\lambda_0
\end{equation}
Moreover, the phase of the bosonic field may be gauged away by
promoting the constraint into a field.
The remaining degree of freedom of the boson is
consequently its amplitude. Having gauged away the phase of
the slave boson brings us to the {\it radial gauge}.

Certainly, the expectation value of any physical observable
has to be gauge invariant. In the conventional Cartesian
representation --- with the integration over both, the real
and imaginary components of the bosonic field --- the particle
number of bosons (holes) is fixed within a gauge-invariant,
conserving approximation \cite{Kroha}. In the radial
representation, however, the gauge is usually fixed in order
to reduce the bosonic field to a real field, and the number of
bosons is fluctuating. Only in the final integration over the
Lagrange multiplier field $\lambda(\tau)$, the particle number
conservation is ensured at each time step. This situation is
somewhat reminiscent of the theoretical modeling of
superconductivity where, in BCS theory, the phase of the
condensate is fixed, but gauge invariance is restored in RPA
\cite{Anderson58}. Nonetheless the situation for
superconductivity is distinct because there, a {\it global} gauge
is broken, whereas here a broken {\it local} gauge has to be
restored.

In a specific gauge, the correctness of the representation can be
verified through the evaluation of the partition sum, and any
correlation function,
in the atomic limit. In this case the calculation is
straightforward. For example, the partition
sum reads: \begin{equation}
Z_{\rm at} = \int_{-\pi/\beta}^{\pi/\beta} \frac{\beta
   d\lambda }{2 \pi} e^{\beta (i\lambda +\lambda_0)}
   Z_f Z_b =1 + 2 e^{\beta \mu}
\end{equation}
where $Z_f = [1+e^{-\beta (i\lambda +\lambda_0-\mu)}]^2$ and
$Z_b = [1-e^{-\beta (i\lambda +\lambda_0)}]^{-1}$. Note that
the projection onto the physical subspace ($\int d\lambda$)
follows from a contour integral in the complex $z$-plane, with
$z \equiv e^{-i \beta \lambda}$, and it only picks up the
residue at $z=0$.

One may try to carry out this program in the polar
representation, namely by introducing the amplitude $r_n$ and
the phase $\varphi_n$ of the bosonic field:
\begin{equation}
\int_{-\infty}^{\infty} \int_{-\infty}^{\infty} \frac{d
   b_n' db_n''}{\pi} \rightarrow \int_{0}^{\infty}
   d(r_n^2)\int_{0}^{2 \pi} \frac{d\varphi_n}{2\pi}
   \quad .
\end{equation}
The bosonic part of the partition sum can be
calculated \cite{Ito}, and the result is again correctly
$Z_b$. Nevertheless, in this way one does not promote
$\lambda$ into a field, and the phases $\varphi_n$ are not gauged
away and, more importantly, the action to lowest order in the time step, 
is not a bilinear form in $r$ and $\varphi$ \cite{Ito}.

We now start to set up the path integral representation in the
``radial gauge''. In the first place we observe that,
contrary to speculations that the radial representation
simply results from a straightforward coordinate
transformation, it needs to be set up on a discretized time
mesh from the beginning. A translation which respects the requirement that
the integrals converge, irrespective of the sequence of
the various integrations, is the following:
\begin{equation}\label{Zat0}
Z_{{\rm at}} =
\lim_{N\rightarrow\infty} \lim_{\epsilon\rightarrow 0^{+}}
\lim_{\Omega\rightarrow\infty}  \int_{0}^{\infty}
D(r_n^2)\int_{-\Omega}^{\Omega} D\left[\frac{\delta\cdot
d\lambda_n}{2 \pi}\right] \int \prod_{\sigma}
   D[f_{n,\sigma},f^{\dagger}_{n,\sigma}] e^{-S_f-
   S_b+\delta\sum_n(i\lambda_n +\lambda_0)}
\end{equation}
where
\begin{eqnarray}\label{Sloc}
S_f& = & \sum_{n=1}^{N}\sum_{\sigma}
f^{\dagger}_{n,\sigma} \left[ f_{n,\sigma} -
f_{n-1,\sigma} \left( 1-\delta(i\lambda_n +\lambda_0-
\mu)\right )\right ]
\nonumber\\
S_b& = & \delta \sum_{n=1}^{N} (r_n^2-\epsilon) (i\lambda_n
+\lambda_0) \end{eqnarray}
and $\delta = \beta/N$ with $N$ the number of time steps.

We proceed with the evaluation of the atomic partition sum,
step by step, in order to convince the reader of the
correctness of this representation: integration over the
Grassmann variables $[f_{\sigma},f^{\dagger}_{\sigma}]$ yields
(we implicitly understand that the limit $N\rightarrow\infty$
has to be taken at the end of the calculation):
\begin{eqnarray}
Z_{\rm at} & = & 
\lim_{\epsilon\rightarrow 0^{+}}
\lim_{\Omega\rightarrow\infty} \left( \prod_{n=1}^{N} \int_{-
\Omega}^{\Omega} \frac{\delta d\lambda_n}{2 \pi} \int_{0}^{\infty}
d(r_n^2) e^{-\delta(i\lambda_n
+\lambda_0)(r_n^2-\epsilon-1)}\right) 
\nonumber\\ & \times & \left[ 1 +\prod_{m=1}^{N}\left(1-\delta(i\lambda_m
+\lambda_0-\mu)\right) \right]^2 \quad , \quad (\Omega>\lambda_0>0)
\end{eqnarray}
The path integral representation, and in particular the
Trotter-Suzuki decomposition, only makes sense when $|\delta
x|\ll 1$, where $x$ denotes anything that multiplies $\delta$
in the course of the calculation. Clearly, a finite
$|\lambda_n|\leq\Omega$ is required in order to reexponentiate 
the fermionic contribution, or, in other words,
to introduce a small parameter, allowing to neglect terms of
order ${\mathcal{O}}(\delta^2)$, for our purpose. The square
bracket in $Z_{\rm at}$ therefore may be written as
\begin{equation}\label{Zfer}
Z_f = \left[ 1 +\prod_{n=1}^{N} e^{-\delta(i\lambda_n
+\lambda_0-\mu)} \right]^2
\end{equation}
and we now obtain
\begin{eqnarray}\label{Zat1}
Z_{\rm at}  &=&  \lim_{\epsilon\rightarrow 0^{+}}
\lim_{\Omega\rightarrow\infty} \left(\prod_{n=1}^{N}
\int_{-\Omega}^{\Omega}
\frac{\delta d\lambda_n}{2 \pi} \int_{0}^{\infty}
d(r_n^2)\right) \nonumber\\
&&\;\left[ e^{-\delta\sum_{n=1}^{N}(i\lambda_n
+\lambda_0)(r_n^2-\epsilon-1)} +
2 e^{\beta \mu} e^{-\delta\sum_{n=1}^{N}(i\lambda_n
+\lambda_0)(r_n^2-\epsilon)}
+ e^{2\beta \mu}e^{-\delta\sum_{n=1}^{N}(i\lambda_n
+\lambda_0)(r_n^2-\epsilon+1)} \right] \nonumber\\
&=&
\lim_{\epsilon\rightarrow 0^{+}}
\lim_{\Omega\rightarrow\infty} \left(\prod_{n=1}^{N} \int_{-
\Omega}^{\Omega}
\frac{\delta d\lambda_n}{2 \pi}\right) \nonumber\\
&&\quad\left[\prod_{n=1}^{N}
\frac{e^{\delta(i\lambda_n
+\lambda_0)(1+\epsilon)}}{\delta(i\lambda_n +\lambda_0)}
+  2 e^{\beta \mu} \prod_{n=1}^{N}
\frac{e^{\delta(i\lambda_n
+\lambda_0)\epsilon}}{\delta(i\lambda_n +\lambda_0)} +
e^{2\beta \mu} \prod_{n=1}^{N} \frac{e^{-\delta(i\lambda_n
+\lambda_0)(1-\epsilon)}}{\delta(i\lambda_n +\lambda_0)}
\right] \quad .
\end{eqnarray}
The origin of $\epsilon$ is now clear: it prescribes how to
close the contour for the second term in Eq.~(\ref{Zat1}). A
motivation, how to find $\epsilon$ and the bosonic Lagrange
density in radial representation, is presented in Appendix~\ref{app:rad}.

In each term, the existence of the integrals is guaranteed by the
shift of the $\lambda$-integration contour into the lower
complex plane induced by a positive $\lambda_0$. The residue
of the pole at $\lambda = i\lambda_0 $ of the contour
integration in the first two terms is finite and yields
exactly the two contributions of
$Z_{\rm at}$: the hole line and the two particle lines
with single occupancy of Fig.~1, respectively. The third term
in Eq.~(\ref{Zat1}), corresponding to double occupancy, is
annihilated through the $\lambda$-integration, as expected
from the implementation of the constraint.

It is instructive to invert the sequence of the integrations.
\begin{eqnarray}\label{Zat2}
Z_{\rm at} & = & \lim_{\epsilon\rightarrow 0^{+}}
\prod_{n=1}^{N} \int_{0}^{\infty} d(r_n^2)
\lim_{\Omega\rightarrow\infty}  \int_{-\Omega}^{\Omega}
\frac{\delta d\lambda_n}{2 \pi}  \nonumber\\
&&\;\left[ e^{-\delta (i\lambda_n
+\lambda_0)(r_n^2-\epsilon-1)}
+ 2 e^{-\delta (i\lambda_n
+\lambda_0)(r_n^2-\epsilon)} e^{\delta \mu} + e^{-\delta
(i\lambda_n
+\lambda_0)(r_n^2-\epsilon+1)} e^{2 \delta \mu}
\right]\nonumber\\ &=& \lim_{\epsilon\rightarrow
0^{+}}\prod_{n=1}^{N} \int_{0}^{\infty} d(r_n^2) \nonumber\\
&&\left[ e^{-\delta \lambda_0 (r_n^2-1)}\hat\delta
(r_n^2-1) + 2 e^{-\delta \lambda_0 (r_n^2-
\epsilon)}\hat\delta (r_n^2-\epsilon) e^{\delta \mu} +
                      e^{-\delta \lambda_0
                      (r_n^2+1)}\hat\delta
(r_n^2+1) e^{2\delta \mu} \right]
\end{eqnarray}
where $\hat\delta$ is the Dirac $\delta$-function.
The last line is equivalent to:
\begin{equation}
Z_{\rm at} = 1 + 2 e^{\beta \mu} \quad . \nonumber
\end{equation}
In this way it appears that, after the projection has been
performed in the first step, $r_n$ may only take two integer
values, $0$ or $1$
in the last line of Eq.~(\ref{Zat2}). We thus arrive at the conclusion
that the integrations over the bosonic
and the constraint fields can be replaced by the simple procedure
demonstrated in Eq.~(\ref{Zat2}).

To summarize, we have introduced a new path integral representation
for slave bosons in the radial gauge. It is defined on a
discretized time mesh, and has a well-defined continuum
limit. Furthermore, we can take a more conventional approach
(Eq.~(\ref{Zat0})), based on a coherent state functional
integral for both fermions and bosons, or use an Ising-like
variable representing the amplitude of empty sites
(Eq.~(\ref{Zat2})). The extension to non-local terms in the
Lagrangian will be discussed in Section~\ref{sec:nonloc}
where we also present a technique how to handle the integration
over the amplitude field $r^2$.

\section{Discrete representation}\label{sec:discrete}

Besides the above integral representation with continuous
fields, which may be used
to set up various approximation schemes, one can
alternatively obtain a discrete representation. Indeed the
above calculation
of the partition sum already suggests that it can be obtained
without having to integrate over continuous values of $r_n$.
Moreover integration over $\lambda_n$ mostly has the effect
of ``picking up'' the appropriate value of $r_n$. This is
achieved more elegantly by extending Coleman's projection
scheme \cite{Coleman} to the case of a time-dependent
constraint. For this purpose one introduces $\xi_n \equiv e^{-
\delta(i\lambda_n + \lambda_0)}$, and rewrites the partition
sum in the following compact form:
\begin{equation}\label{zat0disc}
Z_{\rm at} = \sum_{\{r_n=0,1\}_n} \left(
\prod_{n=1}^{N}\frac{\partial}{\partial \xi_n} \xi_n^{r_n}
\right) \left. \int \prod_{\sigma}
D[f_{\sigma},f^{\dagger}_{\sigma}] e^{-S_f}
\right|_{\xi_1=\ldots = \xi_N=0}
\end{equation}
with
\begin{equation}
S_f =  \sum_{n=1}^{N}\sum_{\sigma}
f^{\dagger}_{n,\sigma}\left[f_{n,\sigma} -
f_{n-1,\sigma}\xi_n e^{\delta\mu}\right]
\end{equation}
In Eq.~(\ref{zat0disc}) and all following relations,
$\sum_{\{r_n=0,1\}_n}$ denotes the ``path integral'' over
the field $r$ which now is reduced to the set of discrete
values $r_n=0,1$ for $n=1,\ldots,N$.
We easily check that the atomic limit is reproduced:
\begin{eqnarray}\label{zat0discres}
Z_{\rm at} &=& \sum_{\{r_n=0,1\}_n} \left. \left(
\prod_{n=1}^{N}\frac{\partial}{\partial \xi_n} \xi_n^{r_n}
\right) \left (1+ \prod_{m=1}^{N}\xi_m e^{\delta\mu}\right)^2
\right|_{\xi_1=\ldots = \xi_N=0} \nonumber\\
&=& 1 + 2 e^{\beta \mu} \quad .
\end{eqnarray}
This discretized representation is certainly the most direct way
to construct the partition sum --- and later on the Green's
function --- as a sum of free fermionic paths, controlled by a
set of Ising variables
$r_1$ to $r_N$.

\subsection{Representation of the Grassmann fields for a
physical electron.}

In the radial gauge we have the following assignment of
electron fields to the previously introduced auxiliary
fields: \begin{eqnarray}\label{CGrass}
c^{\dagger}_{n,\sigma} &=& r_n f^{\dagger}_{n,\sigma}
\nonumber\\ c_{n,\sigma} &=& r_{n+1} f_{n,\sigma}
\end{eqnarray}
The product composition is just the expression
Eq.~(\ref{Cop}) --- only the choice of the time steps has to
be confirmed. This is easily derived from the requirement
that in order to attain a non-zero result for the Green's
function we need ``complete chains'' in the Grassmann
integration. We first clarify the term ``complete chain'' in
the following paragraph, and then convince ourselves
in Subsection~\ref{sec:GF}
that the choice of time steps in Eq.~(\ref{CGrass}) is necessary
to obtain a finite electron propagator.

``Complete chain'' denotes a product of
Grassmann numbers over all time steps, either of
the form $\prod_{n=1}^{N}f_{n,\sigma}
f^{\dagger}_{n,\sigma}$ or, equivalently, as
$\prod_{n=1}^{N}f^{\dagger}_{n,\sigma} f_{n-
1,\sigma}$
which both result from the expansion of the exponentiated action,
and only the integration of such complete chains results in a 
non-zero value of the path integral.
The first form can be characterized as a ``hole chain'' since it
corresponds to the hole line in Fig.~1 (top line). Actually, this
hole contribution
to $Z_{\rm at}$ with value 1 is formed by the product of two
hole chains $(\prod_{n=1}^{N}f_{n,\uparrow}
f^{\dagger}_{n,\uparrow}) (\prod_{n=1}^{N}f_{n,\downarrow}
f^{\dagger}_{n,\downarrow})$.
In the presented slave boson scheme, this double hole chain is
matched by a complete chain of $r_n$ variables with value 1 (the
first contribution in the last line of Eq.~(\ref{Zat2})). The
second arrangement $\prod_{n=1}^{N}f^{\dagger}_{n,\sigma} f_{n-
1,\sigma}$
then should be interpreted as a ``particle chain'' which carries
spin $\sigma$. Indeed, the dashed lines with value
$\exp(\mu\beta)$ in Fig.~1 are formed by
$(\prod_{n=1}^{N}f_{n,-\sigma} f^{\dagger}_{n,-\sigma})
(\prod_{n=1}^{N}f^{\dagger}_{n,\sigma} f_{n-
1,\sigma}e^{\delta\mu})$ with $\sigma=\uparrow,\downarrow$ for
the two lines, respectively. Finally, the third contribution in
the last line of Eq.~(\ref{Zat2})
is generated from a double particle line
$(\prod_{n=1}^{N}f^{\dagger}_{n,-\sigma} f_{n-1,-
\sigma}e^{\delta\mu}) (\prod_{n=1}^{N}f^{\dagger}_{n,\sigma} f_{n-
1,\sigma}e^{\delta\mu})$
but it is projected out via $\hat\delta(r_n^2 + 1)$, as we work in
the constrained Fock space with no double occupancies.

\subsection{Calculation of the Green's function.}\label{sec:GF}
For the Green's function $G_{\sigma}(\tau_f - \tau_i)$ --- with
electron creation at time step $\tau_i$ and annihilation at
$\tau_f$ --- the complete chains
again have to exist. However they are
built from a distinct arrangement of Grassmann variables from the
Lagrangian and the physical electron creation and annihilation
operators.
In this case, the hole chain runs only from time step 0 to the time
step when the $\sigma$-electron is created (visualized in Fig.~2,
where the electron creation is at time step 1). The $\sigma$-hole
chain is
then replaced by a $\sigma$-particle chain in between time steps 1
and $m$. This product of hole- and particle chains still results in
a finite value for the Grassmann integrations since the electron
creation and annihilation operators at time steps 1 and $m$ provide
the missing Grassmann numbers in order to produce a ``complete
chain'' from 0 to $N$.

In order to verify this arrangement and to understand the shift of
the time arguments by one time step in the relation for
$c_{n,\sigma}$ in Eq.(\ref{CGrass}), we explicitly carry out the
evaluation of the
Green's function
\begin{equation} \label{ZGreen}
-ZG_{\sigma}(m-1) 
\equiv
\langle c_{m
,\sigma} c^{\dagger}_{1,\sigma} \rangle
=
\langle  r_{m+1} f_{m ,\sigma} r_{1}
f^{\dagger}_{1,\sigma}\rangle \end{equation} 
in the atomic limit.
Obviously the $\sigma$-fermion exists for $m-1$ time steps.
When we integrate out the Grassmann fields as above for $Z_{\rm at}$
we have: 
\begin{equation}\label{ZGreenat}
-Z_{\rm at}G_{{\rm at},\sigma}(m-1) = \sum_{\{r_n=0,1\}_n} \left. \left(
\prod_{n=1}^{N}\frac{\partial}{\partial \xi_n} \xi_n^{r_n}
\right) r_{m+1} r_{1} \left (1+ \prod_{m=1}^{N}\xi_m
e^{\delta\mu}\right)^2 D^{-1}_{m,1} \right |_{\xi_1=\ldots =
\xi_N=0} \end{equation}
Here the integration over the $-\sigma$-fermion produces one
factor $(1+ \prod_{m=1}^{N}\xi_m e^{\delta\mu}) \equiv
\det{[D]}$, and the integration over the $\sigma$-fermion
yields the propagator matrix-element $D^{-1}_{m,1}$
multiplied by the second factor $\det{[D]}$. This propagator
is:
\begin{equation}\label{Propf}
D^{-1}_{m,1} \equiv \prod_{i=2}^{m} \left( \xi_i e^{\delta\mu} \right)
/\det{[D]} \end{equation}
Next we calculate the derivatives at each time step and obtain
\begin{equation}\label{Gat}
-Z_{\rm at}G_{{\rm at},\sigma}(m-1) = \sum_{\{r_n=0,1\}_n} 
r_{m+1} r_{1} \,(\delta_{r_1,1}
\delta_{r_2,0} \ldots \delta_{r_m,0} \delta_{r_{m+1},1} \ldots
\delta_{r_N,1})\, e^{\delta\mu(m-1)}
\end{equation}
As anticipated above, we produced a chain of $r$-values, with 
$r_i=0$ for $2\leq i\leq m$. We find the correct
expression for this projected atomic Green's function
\begin{equation}
-Z_{\rm at}G_{{\rm at},\sigma}(m-1) = e^{\mu(m-1)\delta} \quad .
\end{equation}
If we had erroneously taken $c_{n,\sigma} = r_{n}f_{n,\sigma}$
in Eqs.~(\ref{CGrass}) and (\ref{ZGreen}), we would 
have to replace $r_{m+1}$ by $r_m$ in
Eq.~(\ref{Gat}), and the result would be incorrectly zero, due
to the $\delta_{r_m,0}$ in the $r-$chain. Higher order
correlation functions can be calculated in a similar way.

We summarize that the expressions for the decomposition of the
$c-$operators (Eq.~(\ref{CGrass})), together with the
expression of the local part of the action (Eq.~(\ref{Sloc}))
and the measure of Eq.~(\ref{Zat0}) form a complete
representation of the slave boson functional integral in the
radial gauge in the limit of infinitely strong local
interaction. In the following chapter we will discuss 
non-local terms in the Lagrangian, using exactly this 
representation.

\section{Non-local Lagrangian}\label{sec:nonloc}

Two types of non-local terms have to be included for a full
discussion of physical models for correlated electrons:
kinetic terms, {\it i.e.\/} hopping of electrons, 
and non-local interactions.

We first consider  kinetic terms of the type
(with $n_{i,\sigma}=c^{\dagger}_{i,\sigma} c_{i,\sigma}$):
\begin{equation}
H_t = \sum_{i,j,\sigma} t_{i,j} (1-n_{i,-\sigma})
c^{\dagger}_{i,\sigma} c_{j,\sigma} (1-n_{j,-\sigma})
\end{equation}
which constrain the hopping of electrons (with hopping
amplitude $t_{i,j}$) to the low energy sector of the
Fock space, {\it i.e.}
doubly occupied sites are not created. In the language of
radial-gauge slave bosons we have
\begin{equation}
S_t = \sum_{n} \sum_{i,j,\sigma} \delta\; t_{i,j}\, r_{i,n+1}
f^{\dagger}_{i,n+1,\sigma} f_{j,n,\sigma} r_{j,n+1}
\end{equation}
using the relations Eq.~(\ref{CGrass}). A formal solution of
a lattice model with arbitrary hopping is possible since the
action is bilinear
in $f^{(\dagger)}_{\sigma}$ and it is found in analogy to the
atomic limit of the previous section. The $\xi-$derivatives,
which enforce the constraint, are now taken at each time step and
each lattice site: \begin{equation}\label{tZla}
Z = \sum_ {\{r_{i,n}=0,1\}_{i,n}}\left. \left(
\prod_{i,n}\frac{\partial}{\partial \xi_{i,n}}
\xi_{i,n}^{r_{i,n}} \right) \det{[D[\xi_{i,n}]]}^2 \right
|_{\xi_{i,1}=\ldots = \xi_{i,N}=0} \end{equation}
with the fermionic inverse propagator matrix
\begin{equation}\label{fprop}
D_{ij,nn'}[\xi_{i,n}] = \hat{\delta}_{i,j} D_{i,nn'}^{(0)} +
\delta\; t_{i,j} r_{j,n+1} r_{i,n+1} \hat{\delta}_{n,n'+1}
\end{equation}
and its local part:
\begin{equation}\label{fprop0}
D_{i,nn'}^{(0)} = \hat{\delta}_{n,n'} - \xi_{i,n}
e^{\delta\mu} \hat{\delta}_{n,n'+1}
\end{equation}
It is understood that the anti-periodicity is taken care
of implicitly. Although the sum over $\{r_{i,n}=0,1\}_{i,n}$ runs
over all possible values of ``hole occupancies'' on each site
and at each time step, ``unphysical paths'' will be projected
out due to the product of $\xi-$derivatives. Hereby,
unphysical paths are the disconnected paths of the variable
$r_{i,n}$ \cite{paths}.
This observation simplifies the calculation
considerably. An explicit example is presented
in Appendix~\ref{app:two},
where the partition sum of the two-site cluster is
calculated.

Secondly we consider non-local interactions. The infinite-$U$
Hubbard model, that we considered so far, is a highly
idealized model which is unlikely to be realized in nature.
In the attempt to set up a lattice model for interacting
electrons, one has to realize that many other non-local
interaction terms may contribute, (for a recent discussion
see Vollhardt {\it et al.\/} \cite{Vollhardt}). On top of the
expected density-density interactions, we also encounter 
spin-exchange terms, correlated hopping terms, four-site terms,
etc.. Here, using our representation, it turns out that
the density-density interaction term,
$S_{\rm nloc} = \sum_{n} \sum_{i,j,\sigma,\sigma'} \delta\; 
V_{i,j} \, n_{i,n,\sigma} \, n_{j,n,\sigma'}$,
can be included without
introducing a fermionic interaction term. Rather, the
contribution to the action may be rewritten in the form:
\begin{equation}\label{nloc}
S_{\rm nloc} = \sum_{n} \sum_{i,j} \delta\; V_{i,j} (1-
r_{i,n}) (1-r_{j,n})
\end{equation}
where we used the identity $r_{i,n}^2 = r_{i,n}$, valid in
the physical subspace. This Gaussian form of the
``interaction term'' has to be contrasted with the canonical
fermionic representation and the
Cartesian representation for slave bosons where the action
remains a true interaction term. In both latter cases, it cannot
be treated in a straightforward fashion. However, here it appears
as additional Boltzmann factors in the expansion of the partition
sum Eq.~(\ref{tZla}).

The last detail that prevents us from taking advantage of
both the radial representation and a conventional propagator
expansion is the integration range for the bosonic field. Indeed it
would be very tempting to extend it from $[0,\infty]$ to $[-
\infty,\infty]$, such
that one could work in momentum space. However such a step is
unlikely
to yield a meaningful answer, since double occupancy is
related to the value $r_n^2=-1$ (see Eq.~(\ref{Zat2})), 
which would then be inside
the integration range. This difficulty may nevertheless be
circumvented by i) introducing a new integration variable
$x_n \equiv r_n^2$, ii) making use of the fact that $r_n^2 =
r_n$ in the physical subspace, iii) adding a potential term $W
x_n (x_n-1) $ that precisely vanishes in the physical
subspace, iv) extending the integration range for $x_n$ to $[-
\infty,\infty]$ and, v) sending $W$ to infinity at the end of
the calculation that would definitely project out any
occupancy different from zero and one. This transforms the
bosonic part of the action $S_b$ in Eq.~(\ref{Sloc}) into:
\begin{equation} \label{Spot}
S_b = \delta \sum_n \left( (i\lambda_n+\lambda_0)(x_n
-1) + W x_n (x_n - 1) \right) \quad .
\end{equation}
We skipped $\epsilon$ here and in the following equations
since $x_n$ now extends to negative values.
Integration over $x_n$ with measure one, now from $-\infty$
to $+\infty$, yields, in  combination with Eq.~(\ref{Zfer}),
the partition sum in the atomic limit:
\begin{eqnarray}\label{Zpot}
Z_{\rm at}  &=&  \lim_{W\rightarrow\infty}
\left(\prod_{n=1}^{N} \int_{-
\infty}^{\infty} \frac{\delta d\lambda_n}{2 \pi}
\sqrt{\frac{\pi}{\delta W}} e^{\delta(i\lambda_n
+\lambda_0)}
e^{\delta W \left( \frac{i\lambda_n +\lambda_0 -
W}{2W}\right)^2} \right) 
\left[1+\prod_{m=1}^{N} e^{-\delta(i\lambda_m
+\lambda_0 -\mu)} \right]^2 \nonumber\\
&=& \lim_{W\rightarrow\infty} \left(1 + 2 e^{\beta \mu} +
e^{2 \beta (\mu-W)}\right)
\end{eqnarray}
As anticipated above double-occupancy is annihilated by
taking the limit $W\rightarrow\infty$ while, in the
opposite limit $W\rightarrow 0$,  {\it i.e.\/} hadn't we
included this potential term
as is customary in slave boson calculations \cite{Arrigoni},
the third contribution could well carry a
substantial weight \cite{double}.
This scheme can be extended to the calculation of the Green's 
function, and in contrast to the above calculation (Eq. (\ref{Zpot})),
$W$ does not enter the result for $Z_{\rm at} G_{{\rm at},\sigma}(\tau)$, 
as explained in Appendix \ref{app:GF}.
Consequently, in a hopping expansion of
the partition sum, $W$ would be related only to those paths where
there is no change in occupancy on one or several sites over the
entire imaginary time interval.

Finally we apply this scheme to lattice models. The measure
of the functional integral is the same as for the single site problem
except that all fields now carry a site index. The potential term, introduced
above to enforce physical values for the amplitude of the slave bosons, may be
implemented as a global term. It is independent of the site index. In summary,
we find for spin-$\frac{1}{2}$ fermions on a lattice,
interacting through an arbitrary Coulomb-like interaction,
which is locally infinitely strong, the following expressions
for the action in the language of radial-gauge
slave bosons:
\begin{equation}
S= S_f + S_b + S_t
\end{equation}
where:
\begin{eqnarray}\label{faction}
S_f&=& \sum_{i,n,\sigma} f^{\dagger}_{i,n,\sigma} \left[
f_{i,n,\sigma} - f_{i,n-1,\sigma} 
e^{-\delta(i\lambda_{i,n} - \mu)}
\right] \nonumber\\
S_b&=& \delta \sum_{i,n} \left( i\lambda_{i,n}(x_{i,n} -1) +
W x_{i,n} (x_{i,n} - 1)  + V_{i,j} (1-x_{i,n}) (1-x_{j,n})
\right) \nonumber\\
S_t &=& \sum_{n} \sum_{i,j,\sigma} \delta\; t_{i,j} \, x_{i,n+1}
f^{\dagger}_{i,n+1,\sigma} f_{j,n,\sigma} x_{j,n+1} \quad ,
\end{eqnarray}
and the partition sum is given by:
\begin{equation}\label{fpartition}
Z = \lim_{W\rightarrow\infty} 
\left( \prod_{i,n}  \int \prod_{\sigma}
D[f_{i,n,\sigma},f^{\dagger}_{i,n,\sigma}] 
\int_{-\infty}^{\infty}
\frac{\delta d \lambda_{i,n}}{2 \pi} \int_{-
\infty}^{\infty} dx_{i,n} \right) e^{-S}
\end{equation}
The measure is now trivial, and the interaction terms included
in $S_b$ are bilinear.
As an example of how this method works for a
lattice problem we solve the Ising chain in Appendix~\ref{app:KR}.

\section{Conclusions}
Models of strongly interacting lattice fermions have been the focus of
many publications on the Mott-Hubbard metal-insulator transition,
itinerant magnetism and strange metallic phases in recent years.
Apart from the thorough analysis of the Anderson impurity case \cite{Kroha},
slave boson approaches have been mainly restricted to saddle
point evaluations.
However low energy spin fluctuations, for example, 
cannot be implemented adequately, as quantum spin fluctuations are only
included as perturbation.  Furthermore, this technique is not suitable
to discuss non-Fermi liquid behavior, as corrections of higher order in
$1/{\cal N}$ have to be included.

This necessitates to further pursue non-perturbative slave boson approaches.
Here we considered the construction of a functional-integral formalism in
the radial gauge which is not restricted to fluctuations around the
${\cal N}=\infty$ 
saddle point. It is defined on a discretized time mesh, and has a well-defined
continuum limit. We restricted our considerations to models in the limit of
large (positive) on-site energy, that is, we work in the constrained Fock
space with no double occupancy.

These models allow, in the radial gauge, to
perform the functional integrals with real bosonic fields. The phase(s) can
be absorbed in the constraint Lagrange multipliers which are thereby
promoted to real (time-dependent) fields. Dynamics of the radial
fields themselves does not exist (as discussed in Appendix~\ref{app:rad}).
They serve to enforce the constraint for each time step by keeping track
of the motion of empty sites, as exemplified for the local problem in
Eqs.~(\ref{Zat2}) and (\ref{Gat}), and for the itinerant problem in
Eq.~(\ref{fprop}). 
This procedure can be extended to other slave boson representations
as, for example, to the Kotliar-Ruckenstein slave bosons, the representation of
which is given for the Ising chain in Appendix~\ref{app:KR} \cite{Ufini}.

Non-local terms in general render the functional integral unsolvable. Yet
Coulomb-like terms with non-local density-density interactions can be
rewritten in bilinear form with radial fields, that is, they represent Gaussian
terms in the functional integral (Eq.~(\ref{nloc}) in
Section~\ref{sec:nonloc}, and also other non-local interactions
may be rewritten in bilinear form, e.g., the Ising spin coupling
within the Kotliar-Ruckenstein representation, see Appendix~\ref{app:KR}). 
This is neither possible in the canonical fermionic representation nor in the
Cartesian representation for slave bosons. These bilinear terms in the
Lagrangian 
generate a finite dispersion for the (radial) slave particles. In
order to keep the bosonic part Gaussian and to allow for a momentum space
representation, one has to extend the range of integration for the square of the
radial field to minus infinity. It can be implemented with the observation that
the functional integration stays correct if we add an additional
global Gaussian term, 
that is, a potential term $W(x_{i,n}-1)x_{i,n}$ where $x_{i,n}$ is the
square of the radial part 
of the slave boson, and send $W\rightarrow\infty$ at the end. This proved to
be a valid procedure to constrain the amplitude of the slave boson to the
physical values 0 and 1, without introducing any additional complications
(see Appendices~\ref{app:KR} and \ref{app:GF}, and the paragraph with 
Eqs.~(\ref{Spot}) and (\ref{Zpot}) in Section~\ref{sec:nonloc}).

In this article we advanced two mechanisms for the solution of a
functional integral 
for constrained electronic problems: either we keep continuous fields which
are originally introduced on a discrete time mesh but have a
well-defined continuum 
limit, or we enforce the constraint in the first step and thereby
reduce the slave 
boson radial field to an Ising-like variable in a discrete
representation. The first 
approach is summarized in the final expressions Eqs.~(\ref{faction})
and (\ref{fpartition}) 
--- and for Kotliar-Ruckenstein slave bosons in Eq.~(\ref{SIsing}) ---
in which the 
bosonic fields are real fields and the bosonic action is of bilinear
form. The second 
approach is discussed in Section~\ref{sec:discrete}, and extended to
itinerant problems in Section~\ref{sec:nonloc}, Eqs.~(\ref{tZla}) and
(\ref{fprop0}), 
which can be taken as a starting point for a numerical evaluation in the strong
coupling limit that does not build on a Hirsch decoupling. 
Furthermore, non-local interactions can be included without the need
of any additional decoupling.

\section{Acknowledgments}

We gratefully thank M.~Dzierzawa and P.~W\"olfle for several
stimulating discussions, and H. Beck and J.-P. Derendinger for
interesting discussions. R.~F.\ is grateful for the
warm hospitality at the Institut f\"ur Theorie der Kondensierten
Materie of Karlsruhe University, and the EKM of Augsburg University
where part of this work has been done.
T.~K.\  greatly enjoyed the hospitality at the Universit\'e de Neuch\^atel.
We acknowledge the financial support by the fonds national suisse de
la recherche scientifique, the BMBF 13N6918/1 and Sonderforschungsbereich 484 of the
Deutsche Forschungsgemeinschaft.

\appendix
\section{}\label{app:rad}

Below we motivate why the radial amplitude field $r_{n}$
has no dynamics and, furthermore, why the limit of
integration is shifted to the negative real axis by an
infinitesimal amount $\epsilon$. We start from $S_b$
Eq.~(\ref{Lcont}) in Cartesian representation.
\begin{equation}\label{dynb}
S_b = \sum_{n} \left[b^{*}_n ( b_{n} - b_{n-1}) + i \lambda
\delta b^{*}_n b_{n-1} \right]
\end{equation}
and substitute $b_{n} = r_{n} e^{i \varphi_{n}}$.

\begin{eqnarray}
S_b &=& \sum_{n} \left[ r_{n} ( r_{n} - r_{n-1}) + r_{n} r_{n-1}
\left( 1- e^{i \delta \frac{\varphi_{n-1} -
\varphi_{n}}{\delta}}(1-i \delta \lambda) \right) \right]
\nonumber\\
&=& \sum_{n} i \delta (\dot{\varphi}_{n} + \lambda) [r_n^2-\delta
r_{n} \dot{r}_{n}] + \delta r_{n} \dot{r}_{n} + {\rm ``further}\; {\rm terms}\;
{\rm in} \; {\mathcal{O}}(\delta^2){\rm {}''} + {\mathcal{O}}(\delta^3)
\end{eqnarray}
In the last line we observe that $ \sum_{n} \delta r_{n} \dot{r}_{n}
\rightarrow \frac{\beta}{2} \int_0^{\beta} d \tau \frac{d r^2}{d
\tau} = 0$ (due to periodic boundary conditions for the bosonic
field). The ``further terms in ${\mathcal{O}}(\delta^2)$'' can all
be neglected with respect to the linear terms. Here we introduced
the time-dependent constraint field $\lambda_n \equiv
\dot{\varphi}_{n} + \lambda$. The included term quadratic in
$\delta$ is special: if $-\delta r_{n} \dot{r}_{n}$ is positive,
it is always a negligible correction to $r_{n}^2$. However, for
$\delta r_{n} \dot{r}_{n} \ge 0$ it may
interpreted as an infinitesimal shift $\epsilon$ of $r_{n}^2$ to
negative values for $r_{n} \rightarrow 0$. Consequently, the
leading term in $S_b$ now reads:
\begin{equation}
S_b = \sum_{n} i \delta \lambda_n (r_{n}^2 - \epsilon) \quad .
\end{equation}
There is no dynamic term of the field $r_{n}$ --- as there is one
for $b_n$ in Eq.~(\ref{dynb}) --- since this amplitude
field is a real field.
On account of the constraint we avoid the
essential difficulty that one encounters when one uses polar
coordinates as discussed by Edwards and Gulyaev \cite{Ito}.
Particularly, the part of the action in polar coordinates, that
corresponds to the Gaussian
part of the action in Cartesian coordinates, cannot be written
in a Gaussian form. This is true even to first order in
$\delta$, and in \cite{Ito} higher order terms in the polar
fields must be kept as well. 

\section{}\label{app:two}
As an illustration of how our discrete representation yields
the partition sum we explicitly calculate it for the two-site
cluster. An instructive ``warm up'' exercise consists in
discretizing the imaginary time axis into six time steps. In
order to evaluate $Z$ Eq.~(\ref{tZla}) we first choose the
physical path:
\begin{eqnarray}
r_1 = r_2 = 1;& \quad r_3 = r_4 = 0;&\quad r_5 = r_6=1 \quad
\mathrm{on \; site \; 1}
\nonumber\\
\tilde{r}_1 = 0;& \quad \tilde{r}_2 =\ldots =\tilde{r}_5=1;& \quad
\tilde{r}_6 =0 \quad \mathrm{on \; site \; 2}
\end{eqnarray}
Here and in the following we use $\tilde{x}_n \equiv x_{i=2, n}$ for
anything on site 2. To built up the inverse propagator matrix we
observe that all $\xi$'s for which $r$ is equal to 1 can be set to
zero, since the derivatives with respect to those $\xi$'s have to be taken
from the prefactor. As a result the inverse  propagator matrix is:
\begin{equation}
\setcounter{MaxMatrixCols}{12}
D = \begin{pmatrix}
1 & \;\;0\;\; & \;\;0\;\; & \;\;0\;\; & \;\;0\;\; & \;\;0\;\; & \;\;0\;\; &
\;\;0\;\; & \;\;0\;\; & \;\;0\;\; & \;\;0\;\; & 0\\
0 & 1 & 0 & 0 & 0 & 0 & -\delta t & 0 & 0 & 0 & 0 & 0\\
0 & -\xi_3 e^{\delta \mu} & 1 & 0 & 0 & 0 & 0 & 0 & 0 & 0 & 0 & 0\\
0 & 0 & -\xi_4 e^{\delta \mu} & 1 & 0 & 0 & 0 & 0 & 0 & 0 & 0 &
0\\
0 & 0 & 0 & 0 & 1 & 0 & 0 & 0 & 0 & -\delta t & 0 & 0\\
0 & 0 & 0 & 0 & 0 & 1 & 0 & 0 & 0 & 0 & 0 & 0\\
0 & 0 & 0 & 0 & 0 & 0 & 1 & 0 & 0 & 0 & 0 & \tilde{\xi}_1
e^{\delta \mu}\\ -\delta t & 0 & 0 & 0 & 0 & 0 & 0 & 1 & 0 & 0 &
0 & 0\\
0 & 0 & 0 & 0 & 0 & 0 & 0 & 0 & 1 & 0 & 0 & 0\\
0 & 0 & 0 & 0 & 0 & 0 & 0 & 0 & 0 & 1 & 0 & 0\\
0 & 0 & 0 & -\delta t & 0 & 0 & 0 & 0 & 0 & 0 & 1 & 0\\
0 & 0 & 0 & 0 & 0 & 0 & 0 & 0 & 0 & 0 & -\tilde{\xi}_6 e^{\delta
\mu} & 1 \end{pmatrix}
\end{equation}
Along this path the hole hops from site 1 to site 2 at time 2,
and
back to site 1 at time 5. Accordingly one would expect that
$t$ only enters $D$ on two entries: $(2,\tilde{1}) $ and $
(\tilde{4},5)$. But it also appears at two other entries.
Nevertheless the latter two entries, which do not correspond
to physical processes, are easily seen not to contribute to
the determinant. The latter reads: \begin{equation}
\det{[D]} = 1 + (\delta t)^2 e^{4 \delta \mu} \xi_3 \xi_4
\tilde{\xi}_1 \tilde{\xi}_6 \quad .
\end{equation}
If we now let the number of time steps to be $N$ and gather
all second
order processes in $t$ (their number being $N(N-1)$), we
obtain their contribution to the partition sum as
\begin{equation}
Z_2 = 2 \left( {\beta} t \right)^2 e^{(N-2) \delta
\mu} \end{equation}
as it should. With some patience one may extend this
procedure to an arbitrary number of hopping processes, and
we checked that they can all be summed up to
\begin{equation}
Z = 1 + 2 e^{\beta \mu} 2 \cosh{(\beta t)}
\end{equation}
which is the correct answer.

\section{}\label{app:KR}
In this appendix we illustrate the method described in
Section~\ref{sec:nonloc} for the Ising chain. To
that aim we make use of the Kotliar and Ruckenstein
representation of the Hubbard model \cite{Kotliar_R} in the limit
$U\rightarrow\infty$. We thus introduce two auxiliary
fermions $f_{\uparrow}$ and $f_{\downarrow}$, and three
auxiliary bosons $e$, representing empty sites, and
$p_{\uparrow}$ and $p_{\downarrow}$ representing singly
occupied sites. On each site $i$ they are subject to three
constraints:
\begin{eqnarray}
e^{\dagger}_i e_i &+& \sum_{\sigma} p^{\dagger}_{i,\sigma}
p_{i,\sigma} = 1 \nonumber\\
p^{\dagger}_{i,\sigma} p_{i,\sigma} &=& f^{\dagger}_{i,\sigma}
f_{i,\sigma} \quad \sigma = \uparrow,\;\downarrow
\end{eqnarray}
which are respectively enforced by three Lagrange multipliers
denoted by $\alpha_i$ and $\lambda_{i,\sigma}$. The phase of all
three bosons can be gauged away
\cite{Fresard_W,Jolicoeur,Bang}, and the three Lagrange
multipliers are promoted to fields. Since, for the Ising
model, we are working at exactly one electron per site, the
$e$-field, representing empty sites, has to be fixed to zero. 
We also could implement this ``half-filling condition'' in the
standard way by introducing a chemical potential for the electrons.
However here, it is more convenient to introduce a Kronecker 
$\delta$-function ${\hat{\delta}}^K_{e_{i,n},0}$ in the measure
(see Eq.~(\ref{ZIsing}) below). Just setting the $e$-field to zero
from the outset would introduce a spurious divergence in the
measure.
The action is thus:
\begin{eqnarray}\label{SIsing}
S&=&\sum_{i,n,\sigma} f^{\dagger}_{i,n,\sigma} \left[
f_{i,n,\sigma} - f_{i,n-1,\sigma} e^{-i \delta
\lambda_{i,n,\sigma}}\right]\nonumber\\
&+& \delta \sum_{i,n,\sigma} \left( W p_{i,n,\sigma} (
p_{i,n,\sigma} -1) - i\lambda_{i,n,\sigma} p_{i,n,\sigma}
\right) + \delta \sum_{i,n} \alpha_{i,n}
\left(\sum_{\sigma}p_{i,n,\sigma} +e_{i,n} -1\right) \nonumber\\ &+&
\delta
\sum_{i,j,n} J_{i,j} (p_{i,n,\uparrow} - p_{i,n,\downarrow})
(p_{j,n,\uparrow} - p_{j,n,\downarrow}) \quad .
\end{eqnarray}
It consists of a fermionic part, the first contribution, and
the remaining ones form the bosonic
part $S_b$. Here, $J_{i,j}$ is the Ising-spin coupling, and $W$ was
introduced in Eq.~(\ref{Spot}) and the preceding discussion.
After the integration over the fermions we obtain the partition sum as:
\begin{eqnarray}\label{ZIsing}
Z&=&  \lim_{W\rightarrow\infty}
 \left\{ \prod_{i} \left[
\prod_{n=1}^N
\int_{-\infty}^{\infty} \frac{\delta d \alpha_{i,n}}{2 \pi}
\int_{-\infty}^{\infty}\prod_{\sigma} \frac{
\delta d\lambda_{i,n,\sigma}}{2 \pi}
\int_{-\infty}^{\infty} \prod_{\sigma}dp_{i,n,\sigma}
\int_{-\infty}^{\infty} de_{i,n}
\;{\hat{\delta}}^K_{e_{i,n},0}
\right]  \right\} \nonumber\\
&\times& \left[ \prod_{i,\sigma}
\left( 1+\prod_{n=1}^N e^{-i \delta
\lambda_{i,n,\sigma}}\right) \right] e^{-S_b} \quad .
\end{eqnarray}
We first integrate over the constraint
fields $\lambda_{\sigma}$, then over $\alpha$ and finally over
$e$. As a result the term with $W$ 
drops out --- since the $\lambda_{\sigma}$-integrations 
already enforce the constraint to the point that 
$p_{i,n,\sigma}$ can take only the discrete values 0 and 1.
Furthermore we observe that $p_{i,n,\sigma}$
loses its time-dependence since the previous integrations resulted
in the two straight ``world lines'' 
$\prod_{n}\hat{\delta}(p_{in\uparrow}) \hat{\delta}(p_{in\downarrow}-1) +
\prod_{n}\hat{\delta}(p_{in\uparrow}-1) \hat{\delta}(p_{in\downarrow})$
for each site $i$. 
Thus, setting $p_{i,\sigma} \equiv
p_{i,1,\sigma}= \ldots =p_{i,N,\sigma}$ we get:
\begin{eqnarray}
Z&=& \left[\prod_{i} \int_{-\infty}^{\infty} dp_{i,\uparrow}
\int_{-\infty}^{\infty} dp_{i,\downarrow} \left(
\hat{\delta}(p_{i,\uparrow}) \hat{\delta}(p_{i,\downarrow}-1) +
\hat{\delta}(p_{i,\uparrow}-1) \hat{\delta}(p_{i,\downarrow})
\right) \right] \nonumber\\
&\times&e^{-\beta \sum_{i,j} J_{i,j} (p_{i,\uparrow} -
p_{i,\downarrow}) (p_{j,\uparrow} - p_{j,\downarrow})}
\end{eqnarray}
which holds for any coupling matrix $J_{i,j}$ and topology. If
we now restrict ourselves to nearest neighbor interaction and a
chain of length $L$ with open boundary conditions, we can
recursively integrate over $p_{1,\uparrow}$,
$p_{1,\downarrow}$, \ldots, $p_{L,\uparrow}$, and
$p_{L,\downarrow}$ to obtain the known result
\begin{equation}
Z = 2 \left( 2 \cosh{(\beta J)} \right)^{L-1} \;.
\end{equation}

\section{}\label{app:GF}
We calculate the Green's function in the atomic limit
using the action Eq.~(\ref{faction}) in which a global constraint
was introduced so that the radial amplitudes can be integrated from 
$-\infty$ to $\infty$. In combination with the expression of
the Green's function Eq.~(\ref{ZGreen}) and the propagator Eq.~(\ref{Propf})
we get:
\begin{eqnarray} 
Z_{\rm at}  G_{{\rm at},\sigma}(m-1)  &=& \lim_{W\rightarrow\infty}
\left( \prod_{n=1}^{N} \int_{-\infty}^{\infty} dx_n
\int_{-
\infty}^{\infty} \frac{\delta d\lambda_n}{2 \pi} e^{-\delta
(i\lambda_n(x_n-1) + W x_n(x_n-1))} \right) x_1 x_{m+1} \nonumber\\
&\times& \left( 1+ e^{-\delta \sum_{i=1}^N(i\lambda_i-\mu)}
\right) e^{-\delta \sum_{i=2}^m(i\lambda_i-\mu)}
\end{eqnarray}
This relation is easily verified either by straightforward integration
over the Grassmann fields or, more directly, by realizing that 
in Eq.~(\ref{ZGreenat}) one has to replace 
$\xi_n\rightarrow e^{-i\delta\lambda_n}$ in the propagator $D^{-1}$
and in $\det D$, and $(\partial \xi_n^{r_n}/\partial \xi_n)$ is replaced
by the bosonic path integral 
$\int_{-
\infty}^{\infty} \frac{\delta d\lambda_n}{2 \pi} e^{-\delta
(i\lambda_n(x_n-1) + W x_n(x_n-1))}$.
Integration over $\lambda$ yields:
\begin{eqnarray} 
&&Z_{\rm at}G_{{\rm at},\sigma}(m-1)  = \lim_{W\rightarrow\infty}
e^{\delta \mu (m-1)} \left( \prod_{n=1}^{N} \int_{-\infty}^{\infty}
dx_n e^{-\delta W x_n(x_n-1)} \right) x_1 x_{m+1} \nonumber\\
&\times & \left[ \hat{\delta}(x_1-1) 
\left(\prod_{n=2}^{m}\hat{\delta}(x_n) \right)
\left(\prod_{n=m+1}^{N}\hat{\delta}(x_n-1) \right) + e^{\beta \mu}
\hat{\delta}(x_1) 
\left(\prod_{n=2}^{m}\hat{\delta}(x_n+1) \right)
\left(\prod_{n=m+1}^{N}\hat{\delta}(x_n) \right) \right] \nonumber\\
& = & e^{\delta \mu (m-1)}
\end{eqnarray} 
The second contribution above, which is related to double occupancy,
cancels due to the additional projection in the external electron
operators, represented by the factors $x_1 x_{m+1}$ in 
Eq.~(\ref{ZGreen}). For this reason $W$ drops out of the calculation of
the Green's function, in contrast to the calculation of the partition sum. 
This holds for the calculation of higher correlation functions as well.

\newpage
\begin{center}
\large{\bf {Figure captions}}
\end{center}

Figure 1: Contributions to the atomic partition sum for infinite
on-site interaction $U$.

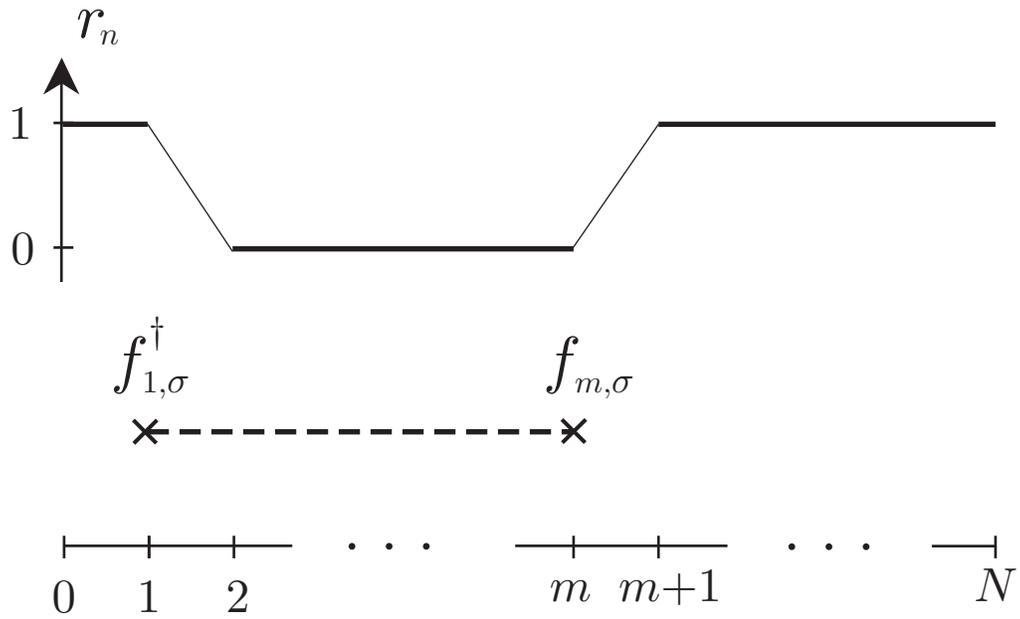
Figure 2: Time evolution of the fields $r$ and
$f_{\sigma}$ in the imaginary-time local Green's
function, $G_{\sigma}(m-1)$. The dashed line is a
``particle chain'' (see text for details). ``Hole
chains'' are not depicted.


\begin{thebibliography}{52}
\bibitem{paths} A connected path is a temporal succession
of $r$-values in which an initial value of $r$ equal to 1 at a
site is continuously ``transported'' through the lattice until the
final time step. For example, a ``transport process'' for which
the value of $r_j$ at site $j$ jumps from 0 at time step $n$ to 1
at time step $(n+1)$ is valid only when it is connected to  ---
via a matrix element $t_{i,j}$ --- a site $i$ where the value of
$r_i$ jumps from 1 at time step $(n+1)$ to 0 at time step $(n+2)$.
For disconnected paths such jumps are not related, and they are
excluded since such paths cannot generate ``complete chains'' (see
previous section).
\bibitem{double} For the atomic limit, this third term in
Eq.~(\ref{Zpot}) with $W\rightarrow 0$ in fact is equal
to the contribution from the state with double occupancy for zero
on-site interaction energy. However,
this observation does not imply that we would have the option to
recover the free limit without on-site interactions for an itinerant
model by applying $W\rightarrow 0$. The phase relation between the
various local states would still be missing. Therefore, the third term
in Eq.~(\ref{Zpot}) is spurious and we always have to take the valid limit
$W\rightarrow\infty$.
\bibitem{Ufini} The original Kotliar and Ruckenstein representation
has been introduced for problems with finite on-site interaction. In
that case an additional field representing double occupancy needs to
be introduced. Direct, but lengthy, calculations show that this field
needs to be complex in order to get the correct result for the
partition sum and correlation functions in the atomic limit.
\end{thebibliography}
\end{document}